\documentstyle[12pt]{article}
\date{Sept 1997/Shahrivar 1376}

\title{Rotation of Plane of Polarization  of an Electromagnetic Wave
Propagating over Cosmological Distances in "Finite Rotating Universe" Scenario}

\author{R.Mansouri\thanks{e\_mail: mansouri@netware2.ipm.ac.ir}
\, and \,
K.Nouzari\thanks{e\_mail: nouzari@netware2.ipm.ac.ir}\\
Cosmology Group, Institute  for Studies in Physics
and Mathematics,\\ P.O.Box 19395-5746, Tehran, Iran.\\
and\\
Department of Physics, Sharif University of Technology,\\ P.O.Box
11365-9161, Tehran, Iran.
}

\begin{document}

\maketitle

\begin{abstract}
A systematic rotation of the plane of polarization of electromagnetic 
radiation propagating over cosmological distances has been claimed to 
be detected by Nodland and Ralston. Taking the finite rotating universe
of Oszvath and Sch\"ucking as a toy model for the actual universe,  
 we calculate the rotation of the plane of polarization of 
electromagnetic waves in this universe. It turns out that the 
observed data are consistent with the Oszvath-Sch\"ucking parameter $k=0.39$.         

\end{abstract}
\newpage
\section{Introduction}
Nodland and Ralstion  have reported a systematic  rotation of  the
plane of polarization of electromagnetic radiation coming from distant
radio galaxis  propagating over
cosmological distances, \cite{1}, \cite{2}. It is claimed that after
extracting the Faraday rotation, there remains an additional rotation 
which may represent evidence for cosmological anisotropy on a vast scale. 
They have found that this  additional rotation follows a dipole rule 
depending on the angular position of the galaxy source and its distance 
to detector. Although this results have been criticized by Cohen
\cite{3}, the possibility of cosmological anisotropy can not yet 
be ruled out. 

There has already been some attempts to explain the results of Nodland and
Ralston. Dobado and Maroto \cite{4} consider the effective coupling between 
electromagnetic and some background torsion field that is obtained
after integrating out charged fermions. Using "Finite Rotating Universe" 
model of Ozsvath and Sch\"ucking we propose here a model description
of the the cosmologically screwy light \cite{2}. 

We first give a brief review of Ozsvath and Sch\"ucking-(O-S)-model 
\cite{5, 6} and then calculate the angle of rotation of 
polarization vector followed by a conclusion. 

\section{Finite Rotating Universe}
The Ozsvath-Sch\"ucking model is a solution of Einstein's field
equations for dust with the following properties:

\begin{enumerate}
\item [1.] The spacetime is the toplogical product $\,R \times S^3\,$. 

\item  [2.] Matter is rotating with non\_zero angular velocity. The 
velocity of rotation with respect to the space or compass of inertia
is given by

\begin{equation}
\label{math:2.1}
\beta = -2 k^2 (\frac{2}{1-2k^2})^{\frac{1}{2}} c
\end{equation}

For $\,k = 0\,$, there follows $\,\beta = 0\,$; that is, for the 
Einstein cosmos the matter is at rest with respect to $S^3$. In the
extreme case $\,k = \frac{1}{2}\,$, the velocity of rotation 
becomes equal to the velocity of light, and the energy\_momentum tensor 
is that of radiation. 

\item [3.] Contrary to the G\"odel cosmos there are no closed 
timelike world lines in this model.
\end{enumerate}
For a complete introduction to O-S model see \cite{6}.

\section{Rotation of the Plane of Polarization of EM Waves and 
the Observational Data}

The rotation in this model is insensitive to the kind of matter. Therefore 
the EM waves are considered to follow the cosmic rotation given by 
 (1). Suppose now an EM wave with a given polarization is propagated 
in the direction of the axis of rotation in the O-S model. After traveling
the cosmological distance x, the plane of polarization of the wave will
be rotated by an amount, $\,\theta\,$, which can be calculated as follows.

Given the velocity of rotation (1), we obtain for the angular velocity 
of rotation

\begin{equation}
\label{math:2.2}
\omega = \frac{\beta}{R} = -\frac{2k^2}{R} (\frac{2}{1-2k^2})^
{\frac{1}{2}} c,
\end{equation}

where $\,R\,$ is the redius of the finite universe. The plane of
polarization of EM wave will rotate by an angular velocity equal to that of
matter. We will therefore have,

\begin{equation}
\label{math:2.3}
x = ct = c \frac{\theta}{\omega}.
\end{equation}

Using equation (\ref{math:2.2}), the angle of rotation is calculated to be

\begin{equation}
\label{math:2.4}
\theta = \frac{\omega x}{c} = - \frac{2k^2}{R} (\frac{2}{1-2k^2})^{\frac{1}{2}}
x
\end{equation}

This gives the amount of twist the light ray is subject to it after 
travelling the distance $x$ along the cosmic axis of rotation.
For $\,k = 0\,$ we obtain $\,\theta = 0\,$ which means that there is 
no rotation of the plane  of polarization in the Einstein cosmos.

Now, taking the values for the residual rotation angle
$\,\beta^{\ast} = 0.4 \ rad\,$, and the travelling distance 
$\,x =  1.5 \times 10^{25} m\,$ from reference \cite{1} and  
$\,R = 2 \times 10^{25} \ m\,$ for the radius of the universe, we obtain 
from the equation (4)

\begin{equation}
\label{math:2.5}
4.5 k^4 + 0.32 k^2 - 0.16 = 0.
\end{equation}

Solving it for $k$ we get

\begin{equation}
\label{math:2.6}
k = 0.3953,
\end{equation}

which lies in the range of $\,0 \leq k \leq \frac{1}{2}$ defined 
by the model. 

\section{Conclusion}

This simplified model shows the possibility of explaining the observational 
data within an exact model based on general relativity. We do not, however, 
 consider the model as an acceptable one, as it has no expansion. It simply
give us an intuition of how such a probable cosmic axis could be 
understood within the standard theories.

\end{document}